\documentclass[aps,prb,superscriptaddress,reprint]{revtex4-1}

\pdfoutput=1

\usepackage{amsmath}
\usepackage{amssymb}
\usepackage{bm}
\usepackage[usenames,dvipsnames]{color}
\usepackage{graphicx}
\usepackage[latin1]{inputenc}
\usepackage[english]{babel}
\usepackage{array}

\begin{document}

\title{All-Optical Study of Tunable Ultrafast Spin Dynamics in [Co/Pd]-NiFe Systems: The Role of Spin-Twist Structure on Gilbert Damping}
%
\author{Chandrima Banerjee}
\author{Semanti Pal}
\affiliation{Department of Condensed Matter Physics and Material Sciences, S. N. Bose National Centre for Basic Sciences, Block JD, Sec. III, Salt Lake, Kolkata 700 098, India}

\author{Martina Ahlberg}
\affiliation{Department of Physics, University of Gothenburg, 412 96, Gothenburg, Sweden}

\author{T. N. Anh Nguyen}
\affiliation{Laboratory of Magnetism and Superconductivity, Institute of Materials Science, Vietnam Academy of Science and Technology, 18 Hoang Quoc Viet, Cau Giay, Hanoi, Vietnam.}
\affiliation{Department of Materials and Nano Physics, School of Information and Communication Technology, KTH Royal Institute of Technology, Electrum 229, SE-16440 Kista, Sweden}

\author{Johan \AA{}kerman}
\affiliation{Department of Physics, University of Gothenburg, 412 96, Gothenburg, Sweden}
\affiliation{Department of Materials and Nano Physics, School of Information and Communication Technology, KTH Royal Institute of Technology, Electrum 229, SE-16440 Kista, Sweden}

\author{Anjan Barman}
\email{abarman@bose.res.in}
\affiliation{Department of Condensed Matter Physics and Material Sciences, S. N. Bose National Centre for Basic Sciences, Block JD, Sec. III, Salt Lake, Kolkata 700 098, India}

\date{\today}

\begin{abstract}

We investigate optically induced ultrafast magnetization dynamics in [Co(0.5 nm)/Pd(1  nm)]$_{5}$/NiFe(\textit{t}) exchange-spring samples with tilted perpendicular magnetic anisotropy using a time-resolved magneto-optical Kerr effect magnetometer. The competition between the out-of-plane anisotropy of the hard layer, the in-plane anisotropy of the soft layer and the applied bias field reorganizes the spins in the soft layer, which are modified further with the variation in \textit{t}. The spin-wave spectrum, the ultrafast demagnetization time, and the extracted damping coefficient all depend on the spin distribution in the soft layer, while the  latter two also depend on the spin-orbit coupling between the Co and Pd layers. The spin-wave spectra change from multimode to single-mode as $t$ increases. At the maximum field reached in this study, $H$=2.5~kOe, the damping shows a nonmonotonic dependence on \textit{t} with a minimum at \textit{t} = 7.5~nm. For \textit{t}~$<$~7.5~nm, intrinsic effects dominate, whereas for \textit{t}~$>$~7.5~nm, extrinsic effects govern the damping mechanisms.

\end{abstract}
\maketitle


\section{Introduction}

Nonuniform magnetic structures, including exchange bias (ferromagnet/antiferromagnet)\cite{Camley1999,Weber2005} and exchange-spring (ferromagnet/ferromagnet)\cite{Haldar2014,Crew2003,Fullerton1998,Fal2011} systems, have recently been explored extensively on account of their intrinsic advantages for applications in both permanent magnets and recording media. Exchange-spring (ES) magnets are systems of exchanged-coupled hard and soft magnetic layers that behave as a single magnet. Here, the high saturation magnetization (\textit{M$_s$}) of the soft phase and the high anisotropy (\textit{H$_k$}) of the hard phase result in a large increase in the maximum energy product. This makes them useful as permanent magnets in energy applications such as engines or generators in miniaturized devices. On the other hand, for spintronic applications, the soft phase is used to improve the writability of the magnetic media, which in turn is stabilized by the magnetic configuration of the hard layer. Consequently, a wealth of research has been devoted to investigating the static and dynamic magnetic properties, including the switching behavior and exchange coupling strength, in ES systems.\\
In case of ES systems with tilted anisotropy, the hard and soft phases consist of materials with out-of-plane (OOP) and in-plane (IP) anisotropies, respectively. This combination results in a canting of the magnetization of the soft layer with a wide and tunable range of tilt angles. The advantage of such a hybrid anisotropy system is that it is neither plagued by the poor writability and thermal instability of systems with IP anisotropy, nor does it lead to very high switching fields, as in OOP systems. As a result, these materials provide additional degrees of freedom to control the magnetization dynamics in magnetic nanostructures, and hint at potential applications in novel spintronic devices utilizing the spin-transfer torque (STT) effect---such as spin-torque oscillators (STOs)\cite{Zhou2008,Zhou2009} and STT-MRAMs.\\
So far, numerous studies have been performed on such systems where the exchange coupling between the hard and soft layers has been tailored by varying the layer thickness,\cite{Nguyen2011,Nguyen2012} layer composition,\cite{Tacchi2013} number of repeats,\cite{Nguyen2014} and interfacial anisotropy.\cite{Nguyen2012} The literature describes investigations of domain structure and other static magnetic properties for [Co/Pd]/Co,\cite{Nguyen-IEEE-2014} [Co/Pd]/NiFe,\cite{Tryputen2015, Nguyen2011, Nguyen-IEEE-2014, Tacchi2013} [Co/Pd]/CoFeB,\cite{Nguyen2014,Tacchi2014, Nguyen-IEEE-2014} [Co/Pd]-Co-Pd-NiFe,\cite{Nguyen2012} [Co/Ni]/NiFe,\cite{Chung2013} and CoCrPt-Ni\cite{Navas2012}---these systems being studied with static magnetometry, magnetic force microscopy (MFM), and micromagnetic simulations. The magnetization dynamics in such systems have also been measured using Brillouin light scattering (BLS)\cite{Tacchi2013,Tacchi2014} and ferromagnetic resonance (FMR)\cite{Tryputen2015} experiments, where the spin-wave (SW) modes have been investigated by varying the thickness of the soft layer and changing the  configuration of the hard layer. In any process involving magnetization dynamics, the Gilbert damping constant (\textit{$\alpha$}) plays a key role in optimizing writing speeds and controlling power consumption. For example, in case of STT-MRAM and magnonic devices, low \textit{$\alpha$} facilitates a lower writing current and the longer propagation of SWs, whereas a higher \textit{$\alpha$} is desirable for increasing the reversal rates and the coherent reversal of magnetic elements, which are required for data storage devices.\\

In this paper, we present  all-optical excitation and detection of magnetization dynamics in [Co(0.5 nm)/Pd(1 nm)]$_{5}$/NiFe(\textit{t}) tilted anisotropy ES systems, with varying soft layer thickness (\textit{t}), using a time-resolved magneto-optical Kerr effect (TR-MOKE) magnetometer. The dynamical magnetic behavior of similar systems has previously been studied using BLS\cite{Tacchi2013} and FMR\cite{Tryputen2015} measurements. However, a detailed study of the precessional magnetization dynamics and relaxation processes in such composite hard/soft systems is yet to be carried out. The advantage of implementing TR-MOKE is that here the magnetization dynamics can be measured on different time scales and the damping is measured directly in the time domain, and is therefore more reliable. We investigate the ultrafast magnetization dynamics over picosecond and picosecond time scales. The ultrafast demagnetization is examined and found to change due to the modified spin structure in the soft layer for different \textit{t} values. The extracted SW spectra are strongly dependent on \textit{t}. An extensive study of the damping coefficient reveals that the extrinsic contribution to the damping is more dominant in the higher thickness regime, while intrinsic mechanisms govern the behavior at lower thicknesses.

\begin{figure}[t!]
\centering
\includegraphics*[width=85mm]{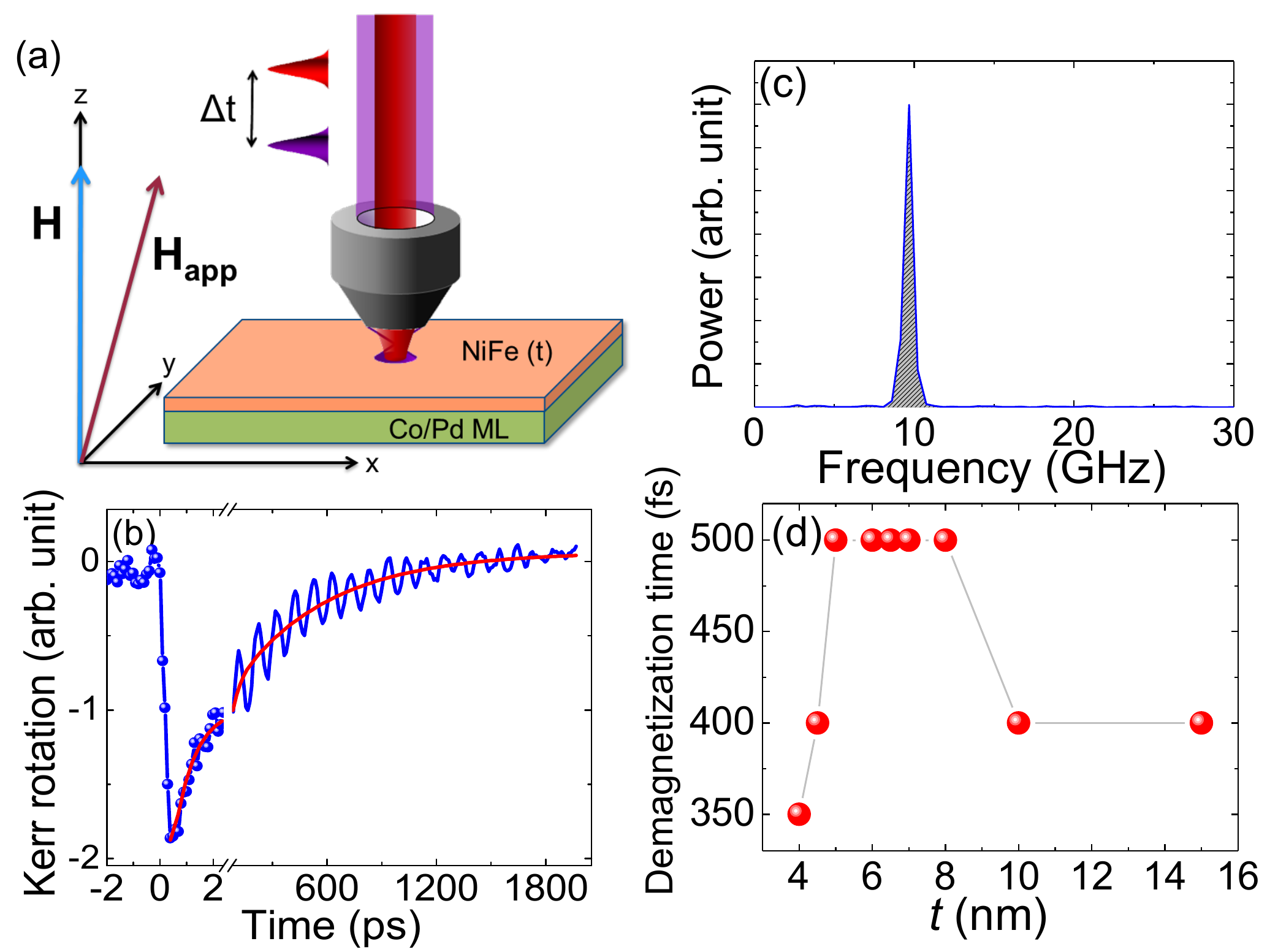}
\caption{(color online) (a) Schematic of the two-color pump-probe measurement of the time-resolved magnetization dynamics of exchange-spring systems. The bias field is applied with a small angle to the normal of the sample plane. (b) Typical time-resolved Kerr rotation data revealing ultrafast demagnetization, fast and slow relaxations, and precession of magnetization for the exchange-spring system with \textit{t} = 7.5~nm at \textit{H} = 2.5~kOe. (c) FFT spectrum of the background-subtracted time-resolved Kerr rotation. (d) Variation of demagnetization time with \textit{t}.}
\label{fig:fig1}
\end{figure}

\section{Experimental details}
\subsection{Sample fabrication}
The samples were fabricated using dc magnetron sputtering and have the following structure: Ta(5nm)/Pd(3nm)/[Co(0.5nm)/Pd(1nm)]$\times 5/\mathrm{Ni}_{80}\mathrm{Fe}_{20}(t)$ /Ta(5nm), where $t=4$--$20$~nm. The chamber base pressure was below $3\times10^{-8}$~Torr, while the Ar work pressure was 2 and 5~mTorr for the Ta, NiFe and Co, Pd layers, respectively. The samples were deposited at room temperature on naturally oxidized Si(100) substrates. The 5~nm Ta seed layer was used to induce fcc-(111) orientation in the Pd layer, which improves the perpendicular magnetic anisotropy of the Co/Pd multilayers; a Ta cap layer was used to avoid oxidation, which has been reported in previous studies.\cite{Nguyen2011, Nguyen2012, Nguyen-IEEE-2014} The layer thicknesses are determined from the deposition time and calibrated deposition rates.

\subsection{Measurement technique}
To investigate the precessional frequency and damping of these samples, the magnetization dynamics were measured by using an all-optical time-resolved magneto-optical Kerr effect (TR-MOKE) magnetometer\cite{Barman-SSP-2014} based on a two-color optical pump-probe experiment. The measurement geometry is shown in Fig.~\ref{fig:fig1}(a). The magnetization dynamics were excited by laser pulses of wavelength (\textit{$\lambda$}) 400~nm (pulse width = 100~fs, repetition rate = 80~MHz) of about 16~mJ/cm$^{2}$ fluence and probed by laser pulses with \textit{$\lambda$} = 800~nm (pulse width = 88~fs, repetition rate = 80~MHz) of about 2~mJ/cm$^{2}$ fluence. The pump and probe beams are focused using the same microscope objective with N.A. of 0.65 in a collinear geometry. The probe beam is tightly focused to a spot of about 800~nm on the sample surface and, as a result, the pump becomes slightly defocused in the same plane to a spot of about 1~$\mu$m. The probe beam is carefully aligned at the centre of the pump beam with slightly larger spot size.  Hence, the dynamic response is probed from a homogeneously excited volume. The bias field was tilted at around~15\textdegree~to the sample normal (and its projection along the sample normal is referred to as \textit{H} in this article) in order to have a finite demagnetizing field along the direction of the pump beam. This field is eventually modified by the pump pulse which induces precessional magnetization dynamics in the samples. The Kerr rotation of the probe beam, back-reflected from the sample surface, is measured by an optical bridge detector using phase sensitive detection techniques, as a function of the time-delay between the pump and probe beams. Figure~\ref{fig:fig1}(b) presents typical time-resolved Kerr rotation data from the ES sample with \textit{t} = 7.5~nm at a bias field \textit{H} = 2.5~kOe. The data shows a fast demagnetization within 500~fs and a fast remagnetization within 8~ps, followed by a slow remagnetization within 1800~ps. The precessional dynamics appear as an oscillatory signal above the slowly decaying part of the time-resolved Kerr rotation data. This part was further analyzed and a fast Fourier transform (FFT) was performed to extract the corresponding SW modes, as presented in Fig.~\ref{fig:fig1}(c).

\begin{figure}[t!]
\centering
\includegraphics*[width=85mm]{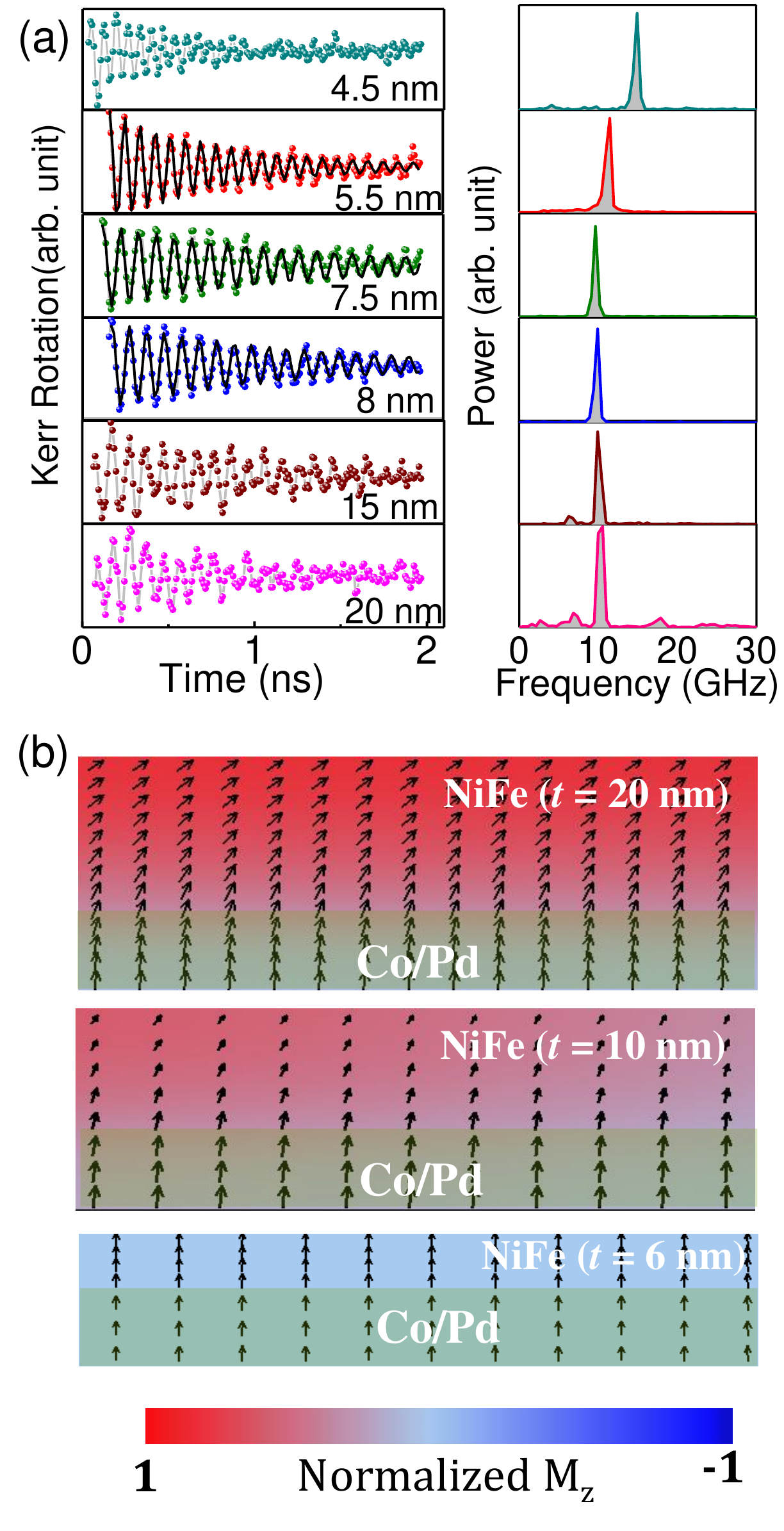}
\caption{(color online) (a) Background-subtracted time-resolved Kerr rotation and the corresponding FFT spectra for samples with different \textit{t} values at \textit{H} = 2.5~kOe. The black lines show the fit according to Eq.~\ref{eq:Mt}. (b) Simulated static magnetic configurations for samples with \textit{t} = 20, 10, and 6~nm with a bias field \textit{H} = 2.5~kOe in the experimental configuration. The simulated samples are not to scale. The color map is shown at the bottom of the figure.}
\label{fig:fig2}
\end{figure}

\section{Results and discussions}
\begin{figure}[t!]
\centering
\includegraphics*[width=85mm]{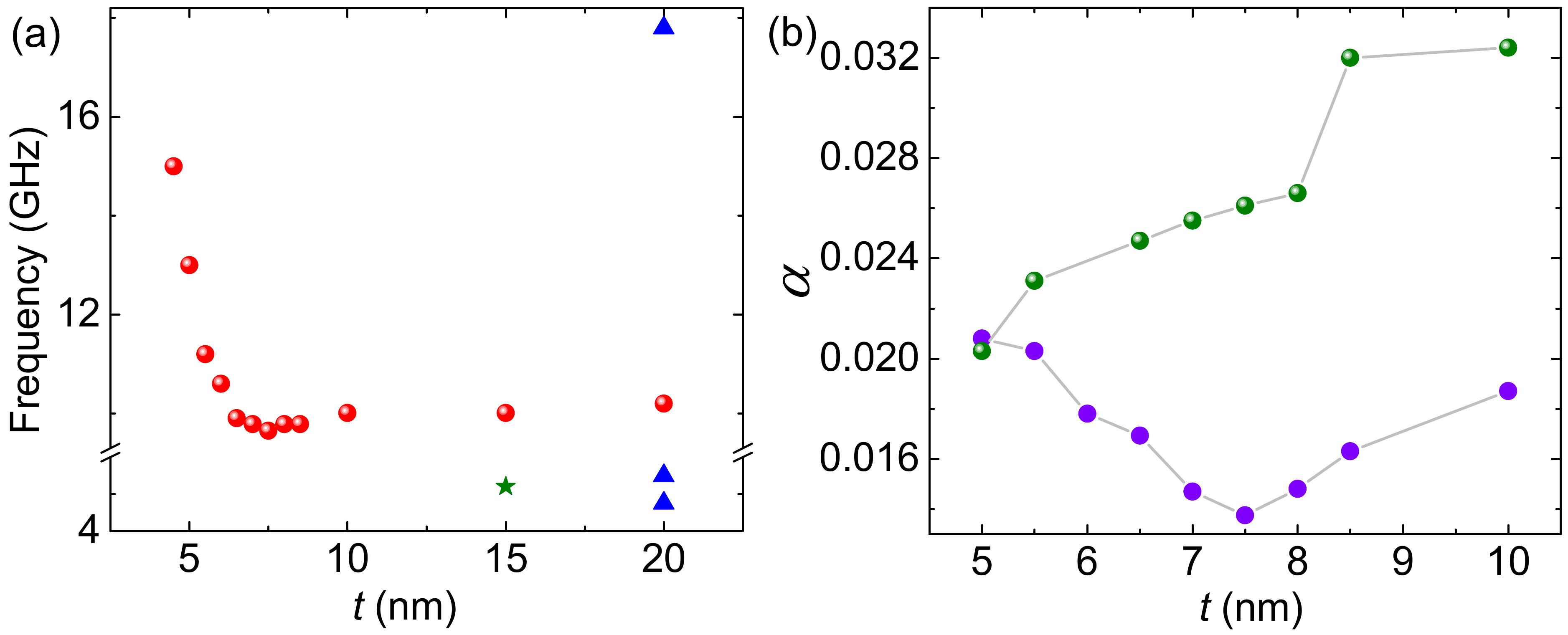}
\caption{(color online) Evolution of (a) spin-wave frequency and (b) Gilbert damping constant as a function of \textit{t} at 1.3~kOe (green circles) and 2.5~kOe (violet circles).}
\label{fig:fig3}
\end{figure}

In order to closely observe the ultrafast demagnetization and fast remagnetization, we recorded the transient MOKE signals for delay times up to 30~ps at a resolution of 50~fs. In Fig.~\ref{fig:fig1}(d), the demagnetization times are plotted as a function of \textit{t}. We observe that the demagnetization is fastest in the thinnest NiFe layer (\textit{t} = 4~nm) and  increases sharply with the increase in \textit{t}, becoming constant at 500~fs at \textit{t} = 5~nm.  At \textit{t} = 10~nm, it decreases drastically to 400~fs and remains constant for further increases in \textit{t}. For \textit{t} $<$ 5~nm, the laser beam penetrates to the Co/Pd layer. In this regime, the large spin-orbit coupling of Pd enhances the spin-flip rate, resulting in a faster demagnetization process. As \textit{t} increases, the top NiFe layer is primarily probed. Here, the spin configuration across the NiFe layer, which is further affected by the competition between the in-plane and the out-of-plane anisotropies of the NiFe and [Co/Pd] layers, governs the demagnetization process. Qualitatively, ultrafast demagnetization can be understood by direct transfer of spin angular momentum between neighboring domains\cite{Vodungbo2012,Koopmans2010}. which may be explained as follows: For \textit{t} $>$ 8~nm, the magnetization orientation in the NiFe layer varies over a wide range of angles across the film thickness, where the magnetization gradually rotates from nearly perpendicular at the Co/Pd and NiFe interface to nearly parallel to the surface plane in the topmost NiFe layer. Such a spin structure across the NiFe layer thickness can be seen as a network of several magnetic sublayers, where the spin orientation in each sublayer deviates from that of the neighboring sublayer. This canted spin structure accelerates the spin-flip scattering between the neighboring sublayers and thus results in a shorter demagnetization time, similar to the work reported by Vodungbo \textit{et al.}\cite{Vodungbo2012} On the other hand, for 5~nm $<$ \textit{t} $<$ 8~nm, the strong out-of-plane anisotropy of the Co/Pd layer forces the magnetization in the NiFe film towards the direction perpendicular to the surface plane, giving rise to a uniform spin structure. The strong coupling reduces the transfer rate of spin angular momentum and causes the demagnetization time to increase.\\
To investigate the variation of the precessional dynamics with \textit{t}, we further recorded the time-resolved data for a maximum duration of 2~ns at a resolution of 10~ps. Figure~\ref{fig:fig2}(a) shows the background-subtracted time-resolved Kerr rotation data for different values of \textit{t} at \textit{H} = 2.5~kOe and the corresponding fast Fourier transform (FFT) power spectra. Four distinct peaks are observed in the power spectrum of \textit{t} = 20~nm, which reduces to two for \textit{t} = 15~nm. This is probably due to the relative decrease in the nonuniformity of the magnetization across the NiFe thickness, which agrees with the variation in the demagnetization time, as described earlier. To confirm this, we simulated the static magnetic configurations of these samples in a field of $H=2.5$~kOe using the LLG micromagnetic simulator.\cite{LLG} Simulations were performed by discretizing the samples in arrays of cuboidal cells with two-dimensional periodic boundary conditions applied within the sample plane. The simulations assume the Co/Pd multilayer as an effective medium\cite{Pal2011} with saturation magnetization \textit{M$_s$} = 690~emu/cc, exchange stiffness constant \textit{A} = 1.3~$\mu$erg/cm, and anisotropy constant \textit{K$_u$$_1$} = 5.8~Merg/cc along the (001) direction, while the material parameters used for the NiFe layer were \textit{M$_s$} = 800~emu/cc, \textit{A} = 1.3~$\mu$erg/cm, and \textit{K$_u$$_1$} =~0.\cite{Pal2014JAP} The interlayer exchange between Co/Pd and NiFe is set to 1.3 $\mu$erg/cm and the gyromagnetic ratio $\gamma$ = 18.1~MHz~Oe$^{-1}$ is used for both layers. The Co/Pd layer was discretized into cells of dimension 5 $\times$ 5 $\times$ 2~nm$^{3}$ and the NiFe layer was discretized into cells of dimension 5 $\times$ 5 $\times$ 1~nm$^{3}$. The results are presented in Fig.~\ref{fig:fig2}(b) for \textit{t} = 20, 10, and 6~nm samples. The nonuniform spin structure is prominent in the NiFe layer of the \textit{t} = 20~nm sample, which modifies the SW spectrum of this sample, giving rise to the new modes.\cite{Pal2014JAP} With the reduction of \textit{t}, the spin structure in the NiFe layer gradually becomes more uniform, while at \textit{t} = 7.5~nm it is  completely uniform over the whole thickness profile. Hence, for low values of \textit{t}, the power spectra shows a single peak due to the collective precession of the whole stack. The variation in precession frequency with \textit{t} is plotted in Fig.~\ref{fig:fig3}(a). The frequency of the most intense mode shows a slow decrease down to \textit{t} = 7.5~nm, below which it increases sharply down to the lowest thickness \textit{t} = 4.5~nm, exhibiting precessional dynamics. This mode is basically the uniform mode of the system and follows  Kittel's equation.\cite{Kittel1948} The variation in frequency depicts the evolution of the effective anisotropy from OOP to IP with increasing \textit{t}, which is in agreement with previously reported results.\cite{Tryputen2015} For lower \textit{t}, the system displays an OOP easy axis, owing to the strong OOP anisotropy of the Co/Pd multilayer. This is manifested as a sharp increase in the frequency with decreasing \textit{t} below 7.5~nm. For greater thicknesses, the effect of the perpendicular anisotropy of the Co/Pd multilayer gradually decreases and the effect of the in-plane NiFe becomes more prominent. These two anisotropies cancel near \textit{t} = 7.5~nm, resulting in a minimum frequency, as shown in Fig.~\ref{fig:fig3}(a).

To extract the damping coefficient, the time domain data was fitted with an exponentially damped harmonic function given by Eq.~\ref{eq:Mt}. 

\begin{equation}
M(t)=M(0)e^{\frac{-t}{\tau}}\sin(2\pi ft-\phi)
\label{eq:Mt}
\end{equation}

\noindent where the relaxation time $\tau$ is related to the Gilbert damping coefficient $\alpha$ by the relation $\tau$ = 1/(2$\pi$\textit{f}$\alpha$). Here, \textit{f} is the experimentally obtained precession frequency and $\phi$ is the initial phase of the oscillation. The fitted data for various values of \textit{t} is shown by the solid black lines in Fig.~\ref{fig:fig2}. We did not extract a value of $\alpha$ for \textit{t} = 15 and 20~nm due to the occurrence of multimode oscillations, which may lead to an erroneous estimate of the damping. The extracted $\alpha$ values are plotted against \textit{t} in Fig.~\ref{fig:fig3}(b) for two different field values of 2.5 and 1.3~kOe. The evolution of $\alpha$ as a function of \textit{t} depends significantly on \textit{H}, as can be seen from the figure. This is because of the different mechanisms responsible for determining the damping in different samples, as will be discussed later. 

\begin{figure}[t!]
\centering
\includegraphics*[width=90mm]{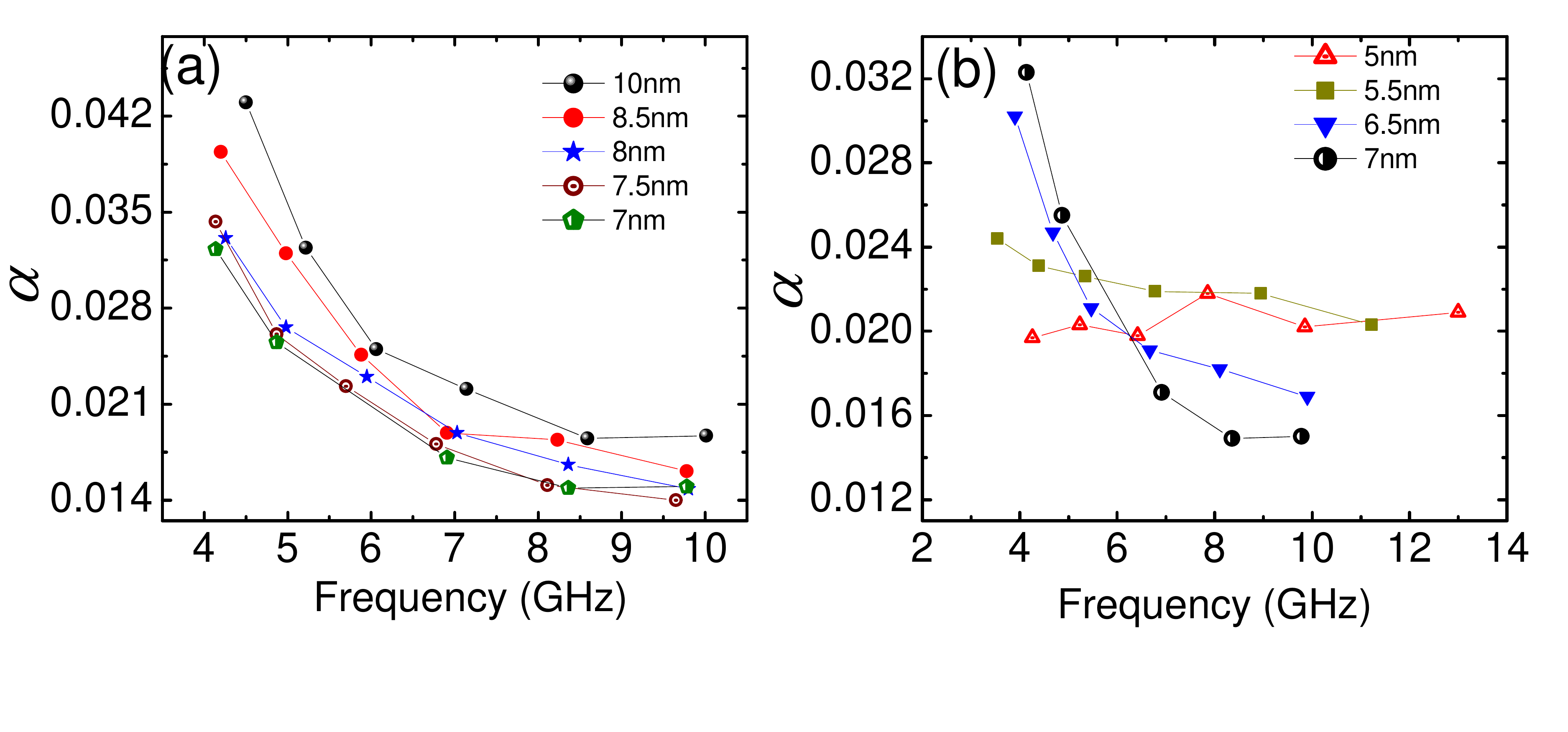}
\caption{(color online) Dependence of Gilbert damping coefficient on soft layer thickness (\textit{t}) for (a) 7--10~nm and (b) 5--7~nm, respectively.}
\label{fig:fig4}
\end{figure}

An interesting trend in the $\alpha$ vs. \textit{t} plot is observed for \textit{H}~=~2.5~kOe. For 10~nm $\leq$ \textit{t} $\leq$ 7.5~nm, $\alpha$ decreases with decreasing \textit{t} and reaches a minimum value of about 0.014 for \textit{t} = 7.5~nm. Below this thickness, $\alpha$ increases monotonically and reaches a value of about 0.022 for the lowest thickness. This variation of $\alpha$ is somewhat correlated with the variation of precession frequency with thickness. In the thinner regime, we probe both the NiFe layer and a fraction of the Co/Pd multilayer and the relative contribution from the latter increases as \textit{t} decreases. The occurrence of a single mode oscillation points towards a collective precession of the stack, which may be considered  a medium with effective magnetic parameters consisting of both NiFe and Co/Pd layers. The variation in damping may be related to the variation in the anisotropy of the material. The competing IP and OOP anisotropies of the NiFe and Co/Pd layers lead to the appearance of a  minimum in the damping. The damping in this system may have multiple contributions, namely (a) dephasing of the uniform mode in the spin-twist structure\cite{Barman2009} (b) interfacial \textit{d}-\textit{d} hybridization at the Co/Pd interface\cite{Pal2011}, and (c) spin pumping into the Pd layer.\cite{Tserkovnyak2002} The first is an extrinsic mechanism and is dominant in samples with higher NiFe thicknesses, while the other two mechanisms are intrinsic damping mechanisms. For \textit{t} $>$ 7.5~nm, due to the nonuniformity of the spin distribution, the dominant mode undergoes dynamic dephasing and the damping thus increasescompared to the magnetically uniform samples. With the increase in NiFe thickness, the nonuniformity of spin distribution and the consequent mode dephasing across its thickness increases, leading to an increase in the damping value. Hence, in samples with higher \textit{t} values, dephasing is the dominant mechanism, while at lower \textit{t} values---i.e., when the contribution from the Co/Pd multilayer is dominant---the spin-orbit coupling and spin pumping effects dominate. At intermediate \textit{t} values, the extrinsic and intrinsic effects compete with each other, leading to a minimum in the damping. However, the damping increases monotonically with \textit{t} in a lower field of \textit{H}=1.3~kOe. For a deeper understanding of this effect, we have measured $\alpha$ as a function of precession frequency \textit{f}. Figures~\ref{fig:fig4}(a)--(b) show the variation of $\alpha$ with \textit{f}. Two different regimes in the thickness are presented in (a) and (b) to show the rate of variation more clearly. For 10~nm $\leq$ \textit{t} $\leq$ 7~nm, $\alpha$ decreases strongly with the decrease in \textit{f} and the rate of variation remains nearly constant with \textit{t}. This is the signature of extrinsic damping generated by the nonuniform spin distribution. However, for \textit{t} = 6.5~nm, the rate falls drastically and for \textit{t} $\leq$~5.5 nm, $\alpha$ becomes nearly independent of \textit{t}, which indicates that purely intrinsic damping is operating in this regime. This confirms the competition between two different types of damping mechanisms in these samples.

The study demonstrates that various aspects of ultrafast magnetization dynamics---namely demagnetization time, precession frequency, number of modes, and damping---are influenced by the spin distribution in the soft magnetic layer, as well as by the properties of the hard layer. By changing the thickness of the soft layer, the relative contributions of these factors can be tuned effectively. This enables efficient control of the damping and other magnetic properties over a broad range, and will hence be very useful for potential applications in spintronic and magnonic devices.

\section{Conclusion}
In summary, we have employed the time-resolved MOKE technique to measure the evolution of ultrafast magnetization dynamics in exchange-coupled [Co/Pd]$_5$/NiFe(\textit{t}) multilayers, with varying NiFe layer thicknesses, by applying an out-of-plane bias magnetic field. The coupling of a high-anisotropy multilayer with a soft layer allows broad control over the spin structure, and consequently other dynamic magnetic properties which are strongly dependent on \textit{t}. The ultrafast demagnetization displayed a strong variation with \textit{t}. The reason for this was ascribed to the chiral-spin-structure-dependent spin-flip scattering in the top NiFe layer, as well as to interfacial 3\textit{d}-4\textit{d} hybridization of Co/Pd layer. The precessional dynamics showed multiple spin-wave modes for \textit{t} = 20~nm and 15~nm, whereas a single spin-wave mode is observed for thinner NiFe layers following the change in the magnetization profile with decreasing \textit{t}. The precession frequency and the damping show strong variation with the thickness of the NiFe layer. The changes in frequency are understood in terms of the modification of the anisotropy of the system, while the variation in damping originates from the competition between intrinsic and extrinsic mechanisms, which are somewhat related to the anisotropy. The observed dynamics will be important for understanding the utilization of tilted anisotropy materials in devices such as spin-transfer torque MRAM and spin-torque nano-oscillators.

\section{ACKNOWLEDGEMENTS}
We acknowledge financial support from the G\"{o}ran Gustafsson Foundation, the Swedish Research Council (VR), the Knut and Alice Wallenberg Foundation (KAW), and the Swedish Foundation for Strategic Research (SSF). This work was also supported by the European Research Council (ERC) under the European Community's Seventh Framework Programme (FP/2007--2013)/ERC Grant 307144 "MUSTANG". AB acknowledges the financial support from the Department of Science and Technology, Government of India (Grant no. SR/NM/NS-09/2011(G)) and S. N. Bose National Centre for Basic Sciences, India (Grant no. SNB/AB/12-13/96). C.B. thanks CSIR for the senior research fellowship.

\end{document}